\documentclass[preprint,12pt,review]{elsarticle}

\usepackage{amssymb}
\usepackage{amsmath}

\begin{document}

\begin{frontmatter}



\title{Short Wavelength Oscillations with Right-Handed Neutrinos}

\author{E.~A.~Paschos\corref{cor1}}

\cortext[cor1]{{\it Email address: {\tt paschos@physik.uni-dortmund.de}}}

\address{Department of Physics, TU Dortmund, D-44221 Dortmund, Germany}

\begin{abstract}
The standard model is extended with three right-handed, singlet neutrinos with general couplings permitted by the $SU(2)_{L}\times U(1)$ symmetry. The traditional oscillations are accounted for, as usually, by three left-handed neutrinos. The article investigates new structures that develop when the masses of the right-handed states are in the eV range. The new states interfere and oscillate with the standard light neutrinos. New structures appear when the detectors average over short wavelengths. I use these results to present and classify properties of the observed  anomalies in the MiniBooNe, reactor and Gallium-detector experiments.
\end{abstract}


\end{frontmatter}

\section{General Introduction}

Among the anomalies in particle physics three were reported in neutrino oscillation experiments. The first appears in the LSND experiment~\cite{Athanassopoulos:1995iw,tba2} which was not confirmed by the MiniBooNe results, but new discrepancies appeared in antineutrino reactions~\cite{AguilarArevalo:2010wv}. The second occurs in reactor experiments where the observed fluxes of neutrinos at a distance of $\sim$100 meters are smaller than those expected from the energy produced by the reactors~\cite{Mueller:2011nm,Mention:2011rk}. Finally, the third anomaly appeared when an intense radioactive source was placed inside the Gallium detectors. These radioactive nuclei decay through electron capture, emitting $\nu_e$ at energies of 1~MeV. The emitted neutrinos have been detected and they show a deficit~\cite{Cribier:2011fv}.

The traditional explanation is the introduction of one or two sterile neutrinos which explain some of the anomalies. The natural way, however, is to introduce one additional right-handed state to each generation and study the consequences. The presence of three states introduces a larger mixing matrix with possible new phenomena. In this article I introduce three right-handed neutrinos -- one for each generation -- with all possible Dirac and Majorana couplings

\begin{equation}
\mathcal{L}_Y=\bar L\,\widetilde{H}_YN_R+\bar N_R M N_R^C+h.~c. \label{eq:1}
\end{equation}

\noindent allowed by the symmetry group $SU(2)_L\times U_1$. The state $L=\left(
\begin{array}{c}
\nu_e\\
e^-\\
\end{array}
\right)_L$ denotes the $SU(2)_L$-left doublets and $\widetilde{H}=i\sigma_2H^*$ the Higgs field. After spontaneous symmetry breaking the neutrino mass matrix $m_D=Yv$ is generated with $v=174$~GeV being the vacuum expectation value of the neutral Higgs. A Majorana mass term is also introduced with $N_{Ri}$'s being $SU(2)_L$ singlets

\begin{equation}
\mathcal{L}_Y^{(\nu)}=\bar\nu_{Li}(m_D)_{ij}N_{Rj}+\frac{1}{2}\bar N_{Ri} M_{ij} N_{Rj}^C+h.~c. \label{eq:2}
\end{equation}

The matrices $m_D$ and $M$ are in general complex.

Using the identity $\bar\nu_{Li}N_{Rj}=\bar N^C_{Rj}\nu^C_{Li}$ half of the Dirac matrix is written as

\begin{equation}
\frac{1}{2}\bar\nu_{Li}(m_D)_{ij}N_{Rj}=\frac{1}{2}\bar N^C_{Rj}(m_D)_{ij}\nu^C_{Li}=\frac{1}{2}\bar N^C_Rm^T_D\nu^C_L \label{eq:3}
\end{equation}

and the entire mass matrix becomes

\begin{equation}
\mathcal{L}_Y^{(\nu)}=\frac{1}{2}\left(\begin{array}{cc}
\bar\nu_L & \bar N_R^C\\
\end{array}
\right)\left(\begin{array}{cc}
0 & m_D\\
m_D^T & M\\
\end{array}
\right)\left(\begin{array}{c}
\nu_L^C\\
N_R\\
\end{array}
\right)+h.~c. \label{eq:4}
\end{equation}

After diagonalization the mass eigenstates are superposition of $\nu_L$ and $N^C_R$ (see equation~\eqref{eq:8} and the discussion that follows). They are  Majorana fermions with the mass matrix including terms like $m_1\bar\phi^C_1\phi_1+m_2\bar\phi^C_2\phi_2+\ldots$ with $\phi_1=\nu_L\cos\theta-N_R^C\sin\theta$ and $\phi_2=\nu_L\sin\theta+N_R^C\cos\theta$, {\it etc.} 

For sterile neutrinos, the bounds on their masses are weak and I will select them to be in the eV mass range. Oscillations to these heavier states influence the detection possibilities. A second criterion deals with the necessary conditions for coherence. One condition requires the neutrino source to be localized within a region smaller than the oscillation length~\cite{Kayser:1981ye,Giunti:1991ca,Cohen:2008qb}. Similarly, the resolution of the detectors must be smaller than the oscillation length; otherwise, we do not observe the undulation of the waves but an average value integrated over several wavelengths.

In the following sections I describe and classify the solutions for the mass matrix in equation~\eqref{eq:4}. It is assumed the $m_D\ll M$ and for simplicity, I restrict the analysis to two flavors: the left-handed states are $\nu_e$ and $\nu_{\mu}$ and the heavier right-handed states will be denoted by $N_e$ and $N_{\mu}$. The mass and energy scales may be different and lead to various cases. We may have the mixing of two states which are light relative to the energy of the beam. This is the traditional case where the observed oscillations have been explained in terms of three light left-handed neutrinos. I will assume that the complete left-handed sector accounts for solar, atmospheric and LBL oscillations. Then I will describe new observables introduced by the right-handed sector. The new observables involve the mixing between light and heavy states and finally the oscillations between two heavy states. In the mixing of light with heavier states the energy difference $E_l-E_h$ can be very large producing a short wavelength, so that an average over several wavelengths is required. We discuss several cases and point out properties that are relevant for experiments. The above properties are present in the case of two or three flavors and for simplicity the analysis is restricted to two flavors where the algebra is simpler and transparent.

It is easy to show that the matrix which diagonalizes the mass matrix is unitary to order $\left(m_D/M\right)^4$, a property that allows one to use standard formulas for appearance and disappearance probabilities. 
The general structure of the mass matrix is dictated by the symmetry properties of the theory without introducing restrictions for the scale of the masses. Choosing a relative small Majorana mass relates the mass matrix to low energy phenomena, but brings the disadvantage that one loses an elegant explanation for the very small neutrino masses. The present experimental results do not rule out small masses and I will present general consequences. 

Sterile neutrinos with relative small masses are popular and there are extensive reviews~\cite{Donini:2012tt,Kopp:2013vaa,Gariazzo:2015rra,Adhikari:2016bei,Abazajian:2012ys}. Among the articles that identify them with Majorana neutrinos there is the so called “eV-seesaw mechanism” with a mass for the right-handed neutrinos in the eV scale~\cite{deGouvea:2005er,deGouvea:2006gz}. This model was introduced to account for the LSND anomaly and was further developed proposing testable predictions~\cite{deGouvea:2009fp,deGouvea:2011zz}. There is also an early review with Dirac and Majorana mass terms with the remark that whenever the resolution of the detectors is large compared to the oscillation length, then the apparatus records an average value of the oscillation~\cite{Zeldovich:1981wf}. 

The plan of the article is a follows. In  section~2. I present a general formalism with extensive details and compute the survival probability for a $\nu_e$ beam. Section 3. describes the appearance probabilities. A classification of the three types of terms in the transition probability is presented in the last section with brief remarks of their implications to other experiments.

\section{Analysis}

We seek solutions in the case where the elements of $m_D$ are small relative to those of $M$ so that perturbative solutions are possible. The unitary matrix which diagonalizes the mass matrix is

\begin{eqnarray}
U=\left(\begin{array}{cc}
(1-\frac{1}{2}JJ^{\dag}) & J\\
-J^{\dag} & (1-\frac{1}{2}J^{\dag}J)\\
\end{array}
\right)\left(\begin{array}{cc}
U_{\theta} & 0\\
0 & U_{\chi}\\
\end{array}
\right)\nonumber\\
= \left(\begin{array}{cc}
(1-\frac{1}{2}JJ^{\dag})U_{\theta} & JU_{\chi}\\
-J^{\dag}U_{\theta} & (1-\frac{1}{2}J^{\dag}J)U_{\chi}\\
\end{array}
\right)\label{eq:5}
\end{eqnarray}

\noindent with $J=m_D\dfrac{1}{M}$ and $U_{\theta}$, $U_{\chi}$ diagonalizing the submatrices $m_D\dfrac{1}{M}m_D^T$ and $M$, respectively. After application of the $U$ matrices the mass matrix is diagonal

\begin{equation}
U^{\dag}\textbf{\Large \textit{m}}\,U^*=\left(\begin{array}{cc}
-U^{\dag}_{\theta}m_D\dfrac{1}{M}m_D^TU_{\theta} & 0\\
0 & U^{\dag}_{\chi}MU_{\chi}\\
\end{array}
\right). \label{eq:6}
\end{equation}

\noindent In this result the accuracy for the $(2,2)$ element is $O\left(m_D\dfrac{1}{M}m_D^T\right)$ and for the $(1,1)$ element $O(m_D^4/M^3)$.

The relations of mass and gauge eigenstates require the inverse matrix, given by

\begin{eqnarray}
U^{-1}=U^{\dag}=\left(\begin{array}{cc}
U_{\theta}^{\dag} & 0\\
0 & U^{\dag}_{\chi}\\
\end{array}
\right)\left(\begin{array}{cc}
(1-\frac{1}{2}JJ^{\dag}) & -J\\
J^{\dag} & (1-\frac{1}{2}J^{\dag}J)\\
\end{array}
\right).\label{eq:7}
\end{eqnarray}

\noindent We define the mass eigenstates $(\phi_1,\phi_2,\phi_3,\phi_4)$ in terms of the flavor states as follows

\begin{equation}
\left[\begin{array}{c}
\phi_1\\
\phi_2\\
\phi_3\\
\phi_4\\
\end{array}\right]=\left(\begin{array}{cc}
U_{\theta}^{\dag} & 0\\
0 & U^{\dag}_{\chi}\\
\end{array}
\right)\left(\begin{array}{cc}
(1-\frac{1}{2}JJ^{\dag}) & -J\\
J^{\dag} & (1-\frac{1}{2}J^{\dag}J)\\
\end{array}
\right)
\left(\begin{array}{c}
\nu_e\\
\nu_{\mu}\\
N_e^C\\
N_{\mu}^C\\
\end{array}\right)\label{eq:8}
\end{equation}

\noindent with the properties we gave for the matrices it is easy to invert these equations and write the flavor states in terms of mass eigenstates and then proceed to compute surviving and appearance probabilities.

For instance we find

\begin{equation}
|\nu_e(t)\rangle=\sum\limits_{i=1}^{4}U_{ei}\phi_i(t)
\label{eq:9}
\end{equation}

with the coefficients given by

\begin{subequations}
\begin{align}
&U_{e1}=[1-\frac{1}{2}(JJ^{\dag})_{11}]c_{\theta}+\frac{1}{2}(JJ^{\dag})_{12}s_{\theta}\label{eq:10a}\\
&U_{e2}=[1-\frac{1}{2}(JJ^{\dag})_{11}]s_{\theta}-\frac{1}{2}(JJ^{\dag})_{12}c_{\theta} \label{eq:10b}\\
&U_{e3}=J_{11}c_{\chi}-J_{12}s_{\chi}\label{eq:10c}\\
&U_{e4}=J_{11}s_{\chi}+J_{12}c_{\chi},\label{eq:10d}
\end{align}
\end{subequations}

and

\begin{subequations}
\begin{align}
&U_{\mu1}=-[1-\frac{1}{2}(JJ^{\dag})_{22}]s_{\theta}-\frac{1}{2}(JJ^{\dag})_{21}c_{\theta}\label{eq:11a}\\
&U_{\mu2}=[1-\frac{1}{2}(JJ^{\dag})_{22}]c_{\theta}-\frac{1}{2}(JJ^{\dag})_{21}s_{\theta} \label{eq:11b}\\
&U_{\mu3}=J_{21}c_{\chi}-J_{22}s_{\chi}\label{eq:11c}\\
&U_{\mu4}=J_{21}s_{\chi}+J_{22}c_{\chi}\label{eq:11d}
\end{align}
\end{subequations}

\noindent with similar equations for the coefficients of the other states. The unitarity of the matrices leads to universality relations, like

\begin{equation}
\sum\limits_{i=1}^{4}|U_{ei}|^2=1+O(J^4)
\label{eq:11}
\end{equation}

\noindent which guarantees that the initial lepton-flavor states have an overall strength determined by the Fermi coupling constant. Unitarity bounds for the matrix element are collected in the review~\cite{Antusch:2014woa}. As the states develop in time the original flavor state diminishes, feeding into the development of states with other flavors.

The experiments produce neutrinos of specific lepton flavor which then develop according to their mass eigenstates. For the time development we can start with equation~\eqref{eq:8} which we invert in order to express $\nu_e(t)$,  $\nu_{\mu}(t)$, $N_e(t)$ and  $N_{\mu}(t)$ in terms of the mass eigenstates $\phi_i(t)$. Then returning to equation~\eqref{eq:8} we can express $\phi_i(0)$'s in terms of $\nu_e(0),\dotsc$,$N_{R\mu}(0)$ being the states produced at time $t=0$. Thus for the development of electron type neutrinos we obtain

\begin{eqnarray}
&|\nu_e(t)\rangle=\left\{\left[\left(1-\frac{1}{2}(JJ^{\dag})_{11}\right)c_{\theta}+\frac{1}{2}(JJ^{\dag})_{12}s_{\theta}\right]^2e^{-iE_1t}\right.&\nonumber\\
&+\left.\left[\left(1-\frac{1}{2}(JJ^{\dag})_{11}\right)s_{\theta}-\frac{1}{2}(JJ^{\dag})_{12}c_{\theta}\right]^2e^{-iE_2t}\right.&\nonumber\\
&+\left.|(J_{11}c_{\chi}-J_{12}s_{\chi})|^2e^{-iE_3t}-|(J_{11}s_{\chi}+J_{12}c_{\chi})|^2e^{-iE_4t}\right\}|\nu_e(0)\rangle+\dotsc\nonumber\\
&=\sum\limits_{i=1}^{4}[C_i]^2e^{-iE_it}|\nu_e(0)\rangle+\dotsc \label{eq:12}
\end{eqnarray}

The ellipses denote terms in the development of $|\nu_e(t)\rangle$ from other initial states. In the limit $J\rightarrow0$ the first two terms that survive have the structure for the two family case. The generalized form includes additional terms from heavier states. The same steps also give the development of a $|\nu_{\mu}(t)\rangle$  from an initial $|\nu_e(0)\rangle$ state

\begin{eqnarray}
|\nu_{\mu}(t)\rangle=\left\{-\left[\left(1-\frac{1}{2}(JJ^{\dag})_{22}\right)s_{\theta}+\frac{1}{2}(JJ^{\dag})_{21}c_{\theta}\right]\right.\nonumber\\
\times\left.\left[\left(1-\frac{1}{2}(JJ^{\dag})_{11}\right)c_{\theta}+\frac{1}{2}(JJ^{\dag})_{12}s_{\theta}\right]e^{-iE_1t}\right.\nonumber\\
+\left.\left[\left(1-\frac{1}{2}(JJ^{\dag})_{22}\right)c_{\theta}-\frac{1}{2}(JJ^{\dag})_{21}s_{\theta}\right]\right.\nonumber\\
\times\left.\left[\left(1-\frac{1}{2}(JJ^{\dag})_{11}\right)s_{\theta}-\frac{1}{2}(JJ^{\dag})_{12}c_{\theta}\right]e^{-iE_2t}\right.\nonumber\\
+\left.(J_{21}c_{\chi}-J_{22}s_{\chi})(J_{11}^*c_{\chi}-J^*_{12}s_{\chi})e^{-iE_3t}\right.\nonumber\\
+\left.(J_{21}s_{\chi}+J_{22}c_{\chi})(J_{11}^*s_{\chi}+J^*_{12}c_{\chi})e^{-iE_4t}\right\}|\nu_e(0)\rangle+\dotsc \label{eq:13}
\end{eqnarray}

Again it is straightforward to compute the ellipses as well as formulas for the development of the two sterile states $|N_{e}(t)\rangle$ and $|N_{\mu}(t)\rangle$. The energies $E_1,\dotsc$,$E_4$ correspond to the mass eigenstates $\phi_1,\dotsc$,$\phi_4$, respectively, and even though $\phi_3$ and $\phi_4$ may be heavier they appear in the $|\nu_{e}(t)\rangle$ and $|\nu_{\mu}(t)\rangle$ wave functions, but as smaller components. The presence of Majorana states influences the oscillation experiments and also neutrinoless double $\beta$-decay.

The contributions of the flavor matrix in equations~\eqref{eq:5} and~\eqref{eq:7} satisfy unitary to $O(J^4)$. This property permits the derivation of general expressions for the transition probabilities similar to those obtained in the traditional cases. 
The transition probability~\cite{Karagiorgi:2006jf,Bilenky:2012zp} that an original neutrino of flavor $\alpha$ to be observed as flavor $\beta$ is

\begin{eqnarray}
P_{\alpha\rightarrow\beta}=\delta_{\alpha\beta}-4Re\sum\limits_{i<j}U^*_{\beta i}U_{\alpha i}U_{\beta j} U^*_{\alpha j}\sin^2\left(\frac{E_i-E_j}{2}t\right)\nonumber\\
+2Im\sum\limits_{i<j}U^*_{\beta i}U_{\alpha i}U_{\beta j} U^*_{\alpha j}\sin\left(E_i-E_j\right)t;\label{eq:14}
\end{eqnarray}

from which the surviving probability for neutrino $\nu_{\alpha}$ follows

\begin{equation}
P_{\alpha\rightarrow\alpha}=1-4\sum\limits_{i<j}|U_{\alpha i}|^2|U_{\beta j}|^2\sin^2\left(\frac{E_i-E_j}{2}t\right).\label{eq:15}
\end{equation}

In these general formulas the mass and energy values are different and it may not be possible to expand the energies in all terms in the relativistic limit. For this reason I kept the energies in the arguments of the trigonometric functions.

For the surviving probability the oscillation between the two light neutrinos has the frequency $\dfrac{\Delta m_{12}^2}{4E}$ which for distances of a few kilometers or shorter produces a small modulation. For the heavier Majorana states, therms $(E_1-E_3)$ and $(E_1-E_4)$ can be large producing an oscillation with short wavelengths which requires the averaging over several wave lengths giving an overall factor $1/2$. The result after averaging over the wavelengths is 

\begin{eqnarray}
P_{\nu_e\rightarrow \nu_e}=1-\left[1-2(JJ^{\dag})_{11}-\frac{2}{\sin 2\theta}Re(JJ^{\dag})_{12}\right]\sin^22\theta\sin^2\frac{\Delta m_{12}^2t}{4E}\nonumber\\
-2\left(|J_{11}|^2+|J_{12}|^2\right)-4|U_{e3}|^2|U_{e4}|^2\sin^2\left(\frac{E_3-E_4}{2}t\right). \label{eq:16}
\end{eqnarray}

There are two corrections of $O(J^2)$ and the last term is $O(J^4)$ which is smaller. The correction to the $\Delta m_{12}-\text{term}$ will be hard to observe, because, as we mentioned, at the short distances $\sin^2\left(\dfrac{\Delta m_{12}^2L}{4E}\right)$ is very small. The term $-2\left(|J_{11}|^2+|J_{12}|^2\right)$ is obtained by averaging over several wavelengths and manifests itself as a decrease of the initial flux  of neutrinos. {\it It is a candidate for the reactor and Gallium discrepancy}. Assigning the decrease of the reactor fluxes to the term $-2\left(|J_{11}|^2+|J_{12}|^2\right)$ implies that one of $|J_{11}|$ or $|J_{12}|$ or both are approximately $\sim0.2$. This range of small parameters will be useful for estimating  the appearance probabilities of $\nu_e$'s in the  MiniBooNe experiments.


The last term in equation~\eqref{eq:16} contains the energy difference $(E_3-E_4)t$. When the masses are different from each other and heavier than the energy of the beam an averaging over wavelengths is again necessary producing an additional reduction of $O(J^4)$. When the masses are comparable $M_3\approx M_4$ the wave length can be long, introducing a modulation on top of the surviving probability. In summary, the presence of heavier neutrinos produces new effects in oscillation experiments. They manifest themselves as a reduction of the surviving probability and/or a modulation on top of the traditional oscillation.

\section{Appearance Probabilities}
For the experiments the relevant parameters are the masses of the neutrinos, the energy of the beam and the mixing angles. A general property is that small mass difference of neutrinos produce oscillations with long wavelengths. For masses and mass differences comparable to the energies the oscillation lengths become small so that the experiments measure rates averaged over several wavelengths.
For active neutrinos there are stringent bounds on their masses but for sterile there is still a large range allowed. For the latter we will consider masses up to 10~eV. The available neutrino beams for reactor experiments are $\approx8$~MeV and for accelerator experiments are a few hundred MeV. In these experiments the neutrino states that we consider are relativistic, so that the functional form

\begin{equation}
(E_i-E_j)L\approx\frac{\Delta m_{ij}^2L}{4E}=1.27\,\frac{\Delta m_{ij}^2}{\text{eV}^2}\,\frac{\text{GeV}}{E_{\nu}}\,\frac{L}{\text{km}}\label{eq:17}
\end{equation}

is justified.



For the experiments~\cite{Athanassopoulos:1995iw,tba2,AguilarArevalo:2010wv,Mueller:2011nm,Mention:2011rk,Cribier:2011fv} there are kinematic regions 
where the wavelength of the oscillation is comparable to the resolution length of the experiments. For $\Delta m^2=1$~eV$^2$ and neutrino energy of $\sim100$ MeV, the distance $L\approx1$~km contains 2 wavelengths and for $\Delta m^2=10$~eV$^2$ contains 20 oscillations. For small mass differences we use the wave nature of the oscillation and for large mass differences one must average over many oscillations. In the following we consider both ranges.

The appearance probability was given in equation~\eqref{eq:14} from which we can distinguish three cases.

Case (1) includes only light neutrinos with $i=1$ and $j=2$ producing the transition

\begin{eqnarray}
P^{(1)}_{\mu\rightarrow e}=-4Re\, U^*_{\mu1}U_{e1}U_{\mu2} U^*_{e2}\sin^2\frac{\Delta_{12}L}{4E}\nonumber\\
\approx\sin^22\theta\sin\frac{\Delta_{12}L}{4E}.\label{eq:18}
\end{eqnarray}

Here $L$ is the distance where the detector is located. For distances less than 100 meters the small value of $\Delta_{12}$ generates very few $\nu_e$ and we kept only the leading term of the mixing matrix which is real.

Case (2) involves contributions from two light neutrinos to heavy states with $i=(1~\text{or}~2)$ and $j=(3~\text{or}~4)$. The masses of the light states are very small and we approximate $\Delta_{ij}\approx-m_j^2=\Delta_j$. The appearance probability is

\begin{eqnarray}
P^{(2)}_{\mu\rightarrow e}=-4Re\left\{\left(U^*_{\mu1}U_{e1}+U^*_{\mu2} U_{e2}\right)\left(U_{\mu3}U^*_{e3}\sin^2\frac{\Delta_{3}L}{4E}+U_{\mu4}U^*_{e4}\sin^2\frac{\Delta_{4}L}{4E}\right)\right\}\nonumber\\
+2Im\left\{\left(U^*_{\mu1}U_{e1}+U^*_{\mu2} U_{e2}\right)\left(U_{\mu3}U^*_{e3}\sin\frac{\Delta_{3}L}{2E}+U_{\mu4}U^*_{e4}\sin\frac{\Delta_{4}L}{2E}\right)\right\}.\label{eq:19}
\end{eqnarray}

In this expression the unitarity of the mixing matrix eliminates the leading terms and leaves the remainder

\begin{equation}
U^*_{\mu1}U_{e1}+U^*_{\mu2} U_{e2}=-\frac{1}{2}\left[(JJ^{\dagger})_{21}+(JJ^{\dagger})_{12}\right]\label{eq:19a}
\end{equation}

which is second order on the parameter $J$. The second multiplicative factor in equation~\eqref{eq:19} is also $O(J^2)$ which makes the effects small. Adopting the value for the element $J_{ij}\sim0.2$ then $4(J_{ij})^4=0.0064$ which is the preferred value by the MiniBooNe Collaboration~\cite{AguilarArevalo:2010wv}. For the mass parameters $\Delta_{13},\ldots\Delta_{24}$ there is a lot of freedom and can be chosen to produce appropriate sinusoidal distribution. Alternatively the masses $m_3$ and $m_4$ can be selected to be heavy enough so that an average over several wavelengths is justified.

When we go a step further and consider the masses $m_3$ and $m_4$ heavy enough so that an average over several wavelengths is justified, then the expression for the appearance probability further simplifies

\begin{equation}
P^{(2)}_{\mu\rightarrow e}=2|U_{\mu3}U^*_{e3}+U_{\mu4}U^*_{e4}|^2=2|J_{21}J_{11}^*+J_{22}J_{12}^*|^2.\label{eq:20}
\end{equation}

The absence of the imaginary term from equation~\eqref{eq:19} follows from the averaging over the length of the detector and also from the fact that the mixing elements appear as an absolute value. Thus for the appearance of $\nu_e$'s there are two possibilities. In the first one the two wavelengths can be chosen appropriately to accommodate experimental data and in the second the wavelengths are very short producing a mean number of~$\nu_e$'s from the oscillation.

Finally case (3) originates from the interference of two heavy intermediate states. This is sensitive to the phase between the matrix elements and includes the CP-phases. I define

$$\phi_{34}=\text{arg}[U_{\mu3}U^*_{e3}U_{\mu4}U^*_{e4}]$$

and obtain the appearance probability

\begin{equation}
P^{(3)}_{\mu\rightarrow e}=|U_{\mu3}||U_{e3}||U_{\mu4}||U_{e4}|\left[-4\cos\phi_{34}\sin^2\frac{\Delta_{34}L}{4E}+2\sin\phi_{34}\sin\frac{\Delta_{34}L}{2E}\right].\label{eq:21}
\end{equation}

The masses $m_3$ and $m_4$ are large but we do not know how close they are to each other. Thus they provide one more wavelength. In general there are three wavelengths  accounting for the data.

\section{Summary}
The article investigated theories with Dirac and Majorana mass terms including all terms allowed by the symmetries of the standard model. We selected low Majorana masses in order to correlate the consequences with low energy phenomena. The results have the following features.

1. The mixing among the light states produces the observed, (atmospheric and solar) oscillations. The standard analysis with three left-handed light neutrinos can be repeated.

2. The mixing of light with the heavy states originates from the middle values in the summations of equations~\eqref{eq:14} and~\eqref{eq:15}. Using the unitarity we can show that they depend on the mixing elements of the heavy states. For short wavelengths they do not include CP-violating effects.

3. At the end of the summation there are two heavier states which interfere with each other. If the mass difference between them is very small then the oscillation has a very long wavelength (LBL). Moderate values for the mass difference produces oscillations accessible to short baseline experiments.
I formulated the discussion for two families but the arguments rely on the unitarity of the mixing matrix and the separation into the three cases applies to models with more families.

The existence of massive neutrinos with masses in the eV region can be investigated in other experiments. There is a group of experiments looking directly for neutrino masses. They investigate the end-point of the Curie-plot which is sensitive to the incoherent sum~\cite{Goswami:2005ng,Riis:2010zm,Formaggio:2011jg}

\begin{equation}
m_{\beta}=\sqrt{\sum\limits_i|U_{ei}|^2m_i^2}
\end{equation}

\noindent where the new masses contribute. A second possibility are contributions to neutrinoless double beta decay~\cite{Rodejohann:2012xd,Pas:2015eia}. In this process the new Majorana states appear as intermediate states but they are lighter than the momentum flowing through the propagator (the momentum in propagator is $\sim100$~MeV). The decay rate now depends on the effective mass~\cite{deGouvea:2005er,Rodejohann:2012xd,Pas:2015eia,Kayser:1984ge}

\begin{equation}
\langle m_{ee}\rangle=\left(U\mathcal{D}U^T\right)_{ee}=0
\end{equation}

\noindent where $\mathcal{D}$ is the diagonal mass matrix in equation~\eqref{eq:6}. In the see-saw mechanism the masses are correlated and give vanishing values. This is a difference  between the eV-seesaw mechanism~\cite{deGouvea:2005er} and models with one or two strile neutrinos~\cite{Donini:2012tt,Kopp:2013vaa,Gariazzo:2015rra,Adhikari:2016bei,Abazajian:2012ys} where the masses  are unconstrained and a small value for $\langle m_{ee}\rangle$ is possible.

Finally, there is a cosmological bound for the sum of the masses of neutrinos which is close to the masses I discussed.

\newpage
{\bf Acknowledgment}

I wish to thank Drs. I.~Alikhanov, I.~Nisandzic, M.~Nowakowski and W.~Rodejohann for helpful discussions and communications.




\end{document}